# Human migration patterns in large scale spatial with the resume data

Qi Nie[a], Jian-Jun Wu[a,*], Xiao-Yong Yan[a,b,*], Jin-Hu Liu[b], Jun Wang[b]

[a]State Key Laboratory of Rail Traffic Control and Safety, Beijing Jiaotong University, Beijing, China

[b]Big Data Research Center, University of Electronic Science and Technology of China, Chengdu, China

*Corresponding authors. Email: jjwu1@bjtu.edu.cn; yanxy@bjtu.edu.cn

**Abstract:** Researches on the human mobility have made great progress in many aspects, but the long-term and long-distance migration behavior is lack of in-depth and extensive research because of the difficult in accessing to household data. In this paper, we use the resume data to discover the human migration behavior on the large scale scope. It is found that the asymmetry in the flow structure which reflects the influence of population competition is caused by the difference of attractiveness among cities. This flow structure can be approximately described by the gravity model of spatial economics. Besides, the value of scaling exponent of distance function in the gravity model is less than the value of short-term travel behavior. It means that, compared with the short-term travel behavior, the long-term human migration behavior is less sensitive. Moreover, the scaling coefficients of each variable in the gravity model are investigated. The result shows that the economic level is a mainly factor on the migration.

## 1. Introduction

Understanding how people move between different cities is not only a basic problem of subjects as economic geography[1], demography[2], traffic science[3,4], but also has important application value in many fields. For example, the short-term human moving behavior analysis between cities (travel behavior analysis) helps us understand the spread of infectious diseases[5] at large spatial scales. Long-term migration behavior analysis between cities also has great significance in the study of city development[6] and the formulation of regional development policy[7]. With the rapid development of modern communication technology and Internet, more and more data can record the mobile footprint of large-scale population, which provides a new opportunity for the study of human mobility. At present, researchers have achieved many results in analyzing the short-term travel behavior of different cities by using detailed phone calls[8,9], social networking platforms sign-in[10,11], GPS record[12] and other data.

Pappalardo *et al*. used mobile phones and GPS data to find two different categories of people - returnees and explorers[13]. Gallotti *et al*. studied the driving trajectories of 780,000 private cars in Italy and found the limitations of a purely descriptive model, then they proposed a stochastic accelerated walk model[14]. Blumenstock *et al*. found that they could infer the social status of users via mobile phones, thus predicting the properties of millions of users[15]. Toole *et al*. argued that personal access patterns were easier to predict through social relations than unfamiliar relationships, and they agreed that geographical factors were an important feature of social relationships[16]. Marguta and Paris applied radiation models to the study of measles outbreaks and found infection cycles in different regions based on human mobility[17]. Xu *et al*. took home as a research point to find the collective mobile mode of individual mobile phone towers[18]. Zhao *et al*. introduced a prototype model with three basic elements to reproduce the dynamic processes of human activity on the line and off the line by taking into account the non-Markovian characteristic of human mobility[19]. Yuan and Raubal used Weibull distribution to simulate the three predefined measurement indices of radius, shape index and entropy, and introduced these into the simulation of human activities[20]. Chen et al. proposed three basic models to predict the moving mode of the next position in human mobility: GMM model to find the global behavior of feasible trajectory, PMM model to simulate the moving trajectory of individual types of the past, the RMM model to gather and simulated moving type in some regions[21]. Li et al. applied the data of human mobility to the SIR model to study information dissemination, and found that step length and radius could control and guide the spread of information[22]. To sum up, researchers have made many useful contributions to the study of human mobility both in data and model.

In fact, traditional data such as household registration which can reflect the population migration between cities is sometimes difficult to obtain for researchers and scholars. That restricts our extensive research into the law of population migration among cities. Recently, researchers made use of database of authors in the scientific papers to study movements between institutes or colleges[23]. In other words, they studied the migration behavior of a particular population between cities. However, due to data restriction and other reasons, we still lack sufficient understanding about the law of universal population migration in the cities.

In this paper, we study large-scale and multi-career population migration pattern among cities by using anonymous job information obtained from a Chinese recruitment website. Data we use is crawled from an API port of the recruitment website. We see the links of start and end points of population migration as networks flow. After we link up all the origin and destination of the migration, flow of the migration forms a huge network. We found that the long-term migration network between different cities have highly asymmetric characteristics. However, the short-term intercity travel network which we form by using micro-blog users' registering data shows the symmetry pattern. It reflects the differentiation of population attraction in different cities as a result of city competition. In order to make clear the flow structure, we introduce the gravity model of spatial economics to describe it approximately. One of the most important concerns for moving behavior is distance. And then we find that the long-term migration's scaling exponent of distance function in the gravity model is greater than the short-term mobility's. Here, the magnitude of the scaling exponent represents the influence degree of the corresponding parameter. So this means the long-term migration behaviors are less sensitive to the space distance than the short-term travel behaviors. To predict the trend of human migration, we have developed an improved gravity model. To do this, firstly, we apply the established gravity model to both the human migration behavior and the human travel behavior. Secondly, the difference between the two indicators of the gravity models in the same location is compared. It is found that the scaling exponent of population migration behavior of GDP indicators is higher than that of travel behavior. However, the scaling exponent of population migration behavior is lower than that of travel behavior. Results suggest that the long-term migration is more attracted by the economy than the short-term human travel, while the population migration is less attracted by the population than the travel behavior. We also find out that human mobility is more attracted by the economy both in migration and travel behavior. This shows that people's travelling is mainly driven by the economic level of cities in long-distance moving behaviors.

## 2. Data and methods
### 2.1. Data
Previous studies of human mobility were mainly based on short-term data, because household data reflecting long-range and long-distance migration behaviors was often difficult to obtain. This limits the depth of the researchers' research into human mobility. In order to complement the research of human mobility, we must find some applicative ways to discuss and study the transfer of large space and long time scale. We analyzed data from more than 1,600,000 anonymous job seekers' moving steps on a job site, bypassing the restrictions of household registration information and studying long-term and long-distance migration behavior from the side. These records of working experience in the resumes contain more than one former workplace (or university). Moreover, they have been in their workplaces(or universities) for a long time(more than one year). Therefore, we can ensure that the experiences of working(or becoming a postgraduate) can represent the migration process of these people. To show the similarities and differences between travel and migration behaviors, we analyzed the Sina micro-blog's check-in data as short-term and long-distance travel behavior which is compared with the migration data represented by resume records. It helps to deepen our understanding of the mechanisms of migration behavior and makes us know what the trends of human moving are.

*2.2. Methods*

*Gravity model*

The primitive gravity model described above is very similar to the law of gravitation in mathematical form. But the original gravity model cannot guarantee that the trip distribution matrix $T$ meets the constraints of the amount of travelling from origin ($O_i = \sum_j T_{ij}$) and the amount of moving from destination ($D_j = \sum_i T_{ij}$). Therefore, researchers in the field of traffic have put forward a double constraints gravity model[24]: $T_{ij} = A_i B_j O_i D_j f(d_{ij})$, where $f(d_{ij})$ is an impedance function which is usually distance, and a power function of generalized trip cost ($d_{ij}^{-\beta}$) or other. $A_i = \dfrac{1}{\sum_j B_j D_j f(d_{ij})}$ and $B_j = \dfrac{1}{\sum_i A_i O_i f(d_{ij})}$ are two equilibrium factors generated in iterative computations. The purpose of the two factors is to make the model predictions satisfy both the constraints of starting and the constraints of attraction.

*Sorensen similarity index*

Sorensen similarity index in biology is used for finding differences of gene sequence[25]. We improved the original Sorensen similarity index in order to compute the similarity of two trip matrices. The calculation formula of Sorensen similarity index is as follows: $SSI = \dfrac{1}{N^2} \sum_i^N \sum_j^N \dfrac{2 \min(T_{ij}', T_{ij})}{T_{ij}' + T_{ij}}$, where $T_{ij}$ is the amount of movement from the starting point $i$ to the destination $j$, and $T_{ij}'$ is the amount of movement calculated by models. Obviously, if the flow calculated by the model is closer to the flow in the actual data, the value of SSI is closer to 1; otherwise, if the difference between the two is greater, the value of SSI is closer to 0.

**3. Results**

## 3.1. Statistical characteristics of human mobility

Statistical analysis of human behavior patterns is the basic way to study the complexity of human behavior. From the temporal and spatial point of view, the research patterns of human mobility can be divided into four categories: short-term scaled traffic of a city, transportation or travel behavior among cities which is of short-term scale, long-term inner-city site selection and long-time scaled migration among cities. In this paper, we focus on travel behavior and human migration between cities.

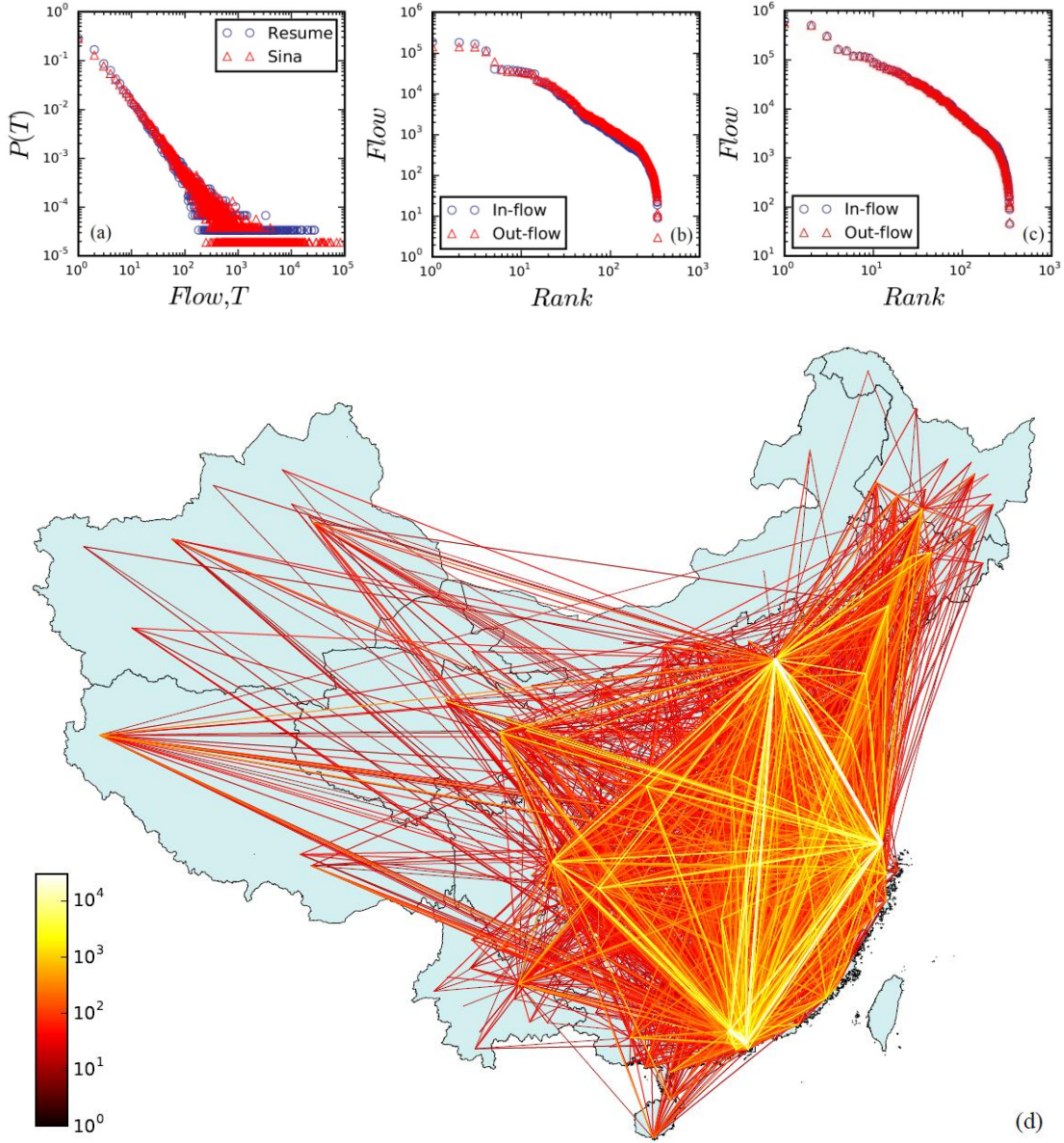

**Fig. 1.** The flow distribution of migration behavior and travel behavior. (a) Both migration behavior (resume) and travel behavior (micro-blog) satisfy the fat tail distribution. (b) Comparison of inflows and outflows based on the rank of cities in migration behavior. The horizontal axis represents the ranking, and the vertical axis indicates the flow. (c) Comparison of inflows and outflows based on the rank of cities in short-term travel behavior. (d) Network of aggregate mobility's flow in migration behavior.

In order to find out the internal mechanism of human migration behavior, we begin with the basic statistical characteristics. Firstly, we visualize the total flow of the migration behavior represented by the resume data and compare migration with the travel behavior represented by the micro-blog check-in data (Fig. 1a). We find out that the overall trend of human movement follows the fat tail distribution. It shows that, from the long-distance movement level, the vast majority of people are used to travelling in a small number of cities. We can see that more than 90% of the cities or regions attract less than 1,000 times of movements. On the contrary, only a small number of cities attract 1,000 or even more than 10,000 of the moving flow. In order to observe the inflow and outflow of migration behavior and travel behavior intuitively, we sort each city according to the amount of flow, and then label the outflow and inflow in the same coordinate system where the horizontal axis represents the city traffic rankings, and the vertical axis represents the actual flow (Fig. 1b and 1c). As we can see, the number of inflows and outflows caused by migration is not equal in the cities where people have more traffic flow (Fig. 1b). Similarly, the same results can be also found in cities with middle ranking which shows a slight imbalance in long-term migration behavior. In the cities with more moving flow, the inflows are obviously larger than the outflows. While in the cities with middle and rear ranking, the inflow and outflow is almost balanced. We can clearly see that inflows of travel behavior almost coincide with outflows whether it's in a big city or in a small one (Fig. 1c). It shows that in short-term travel behavior, people's flow can be considered symmetric and balanced. Meanwhile, in order to visually see the flow of the population in the migration process, we draw the flow of the migration on the map (Fig. 1d). As can be seen, there are strong heterogeneity in migration behavior over long timescales and large spatial scales. In the migration behavior, a large number of people mainly move between Beijing, Shanghai, Guangzhou and other large cities while only a small number of people gather in other small and medium-sized cities. In the figure, the thicker the line represented connection in the map is, the greater the amount of migration between the two places exists; the brighter the line, the more the migration flow.

*3.2. Symmetry analysis*

In fact, it is difficult to observe the symmetry in human mobility from Figure 1. To better illustrate the imbalance of migration behavior, we use the equilibrium figure of human mobility to visually reveal this property (Fig. 2). The abscissa in the figure represents the outflow. The ordinate represents the inflow. The dashed line is the equilibrium line. The colorbar on the right shows the population scale. The larger the circle in the picture is, the more people there are.

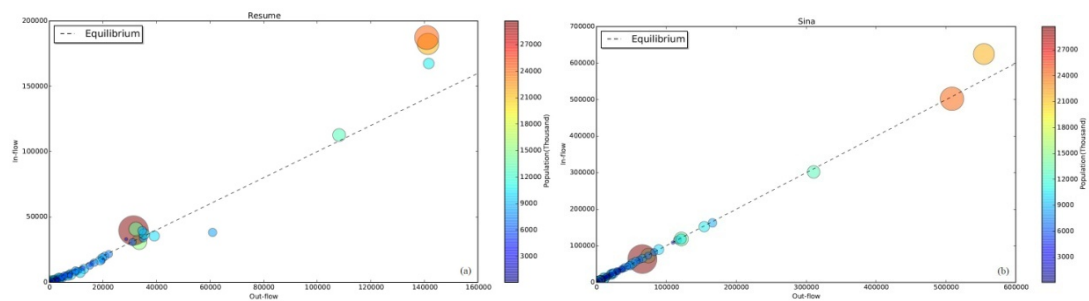

**Fig. 2** Equilibrium diagram of inflows and outflows. Each circle represents a city drawn in the picture according to inflows and outflows. Obviously, the nearer to the equilibrium line (dotted line) the circle is, the closer to the value of outflow the value of inflow is. (a) Equilibrium diagram of migration behavior. The top right cities are Shanghai, Beijing and Shenzhen. The city off the line is Dongguan. The biggest circle represents Chongqing because of the largest population.(b) Equilibrium diagram of travel behavior.

As can be seen from the figure above, the imbalance of the flow (the amount of inflow and outflow is not equal) is very obvious in the migration behavior (Fig. 2a). Shanghai, Beijing and Shenzhen are more attractive for job seekers. This caused many job seekers to move to these three places so that inflows of the three cities are much larger than outflows. The circle representing Dongguan is below the equilibrium line because Dongguan's appeal is relatively smaller than Shenzhen's and Guangzhou's rather than the city is unattractive to job seekers. Dongguan is geographically close to Shenzhen and Guangzhou so that people used to switch their jobs in consideration of better resources. In fact, there are a lot of small cities below the equilibrium line in the lower left corner of figure 2a. This is the reason why Shanghai, Beijing and Shenzhen are above the equilibrium line. In other words, the brain drain in small cities is serious, and these talents tend to go to large cities for work. People used to go home after several trips in their short-term travel behavior (Fig. 2b). So the circles in the picture are close to the equilibrium line; that is to say, inflows and outflows are almost equal in the travel behavior. The inflow of Beijing is obviously larger than the outflow. This is because Beijing, as a cultural center and the capital, attracts a large number of nonlocal tourists in every holiday season and lots of employees of some international groups stationed in Beijing for a long time. Micro-blog's attendance record is on a one-year basis, during which many foreign personnel may be ready to be stationed in Beijing for a long time.

*3.3. Gravity model and distance parameter*

Gravity model is one of the first prediction models of population movement. A long time ago, researchers used the classical gravity model to do a lot of work on human mobility. The basic assumption of gravity model is that the amount of movement between the two places is directly proportional to the population of the two places and inversely proportional to the power function of the distance between the two places[26]:

$$T_{ij} = \alpha \frac{m_i m_j}{d_{ij}^{\beta}} \quad (1)$$

where $T_{ij}$ is the amount of movement from origin $i$ to destination $j$, $m_i$ and $m_j$ are the amount of population at origin $i$ and destination $j$, $d_{ij}$ is the distance between the two places, $\alpha$ and $\beta$ are two adjustable parameters which need to be estimated based on actual data[27].

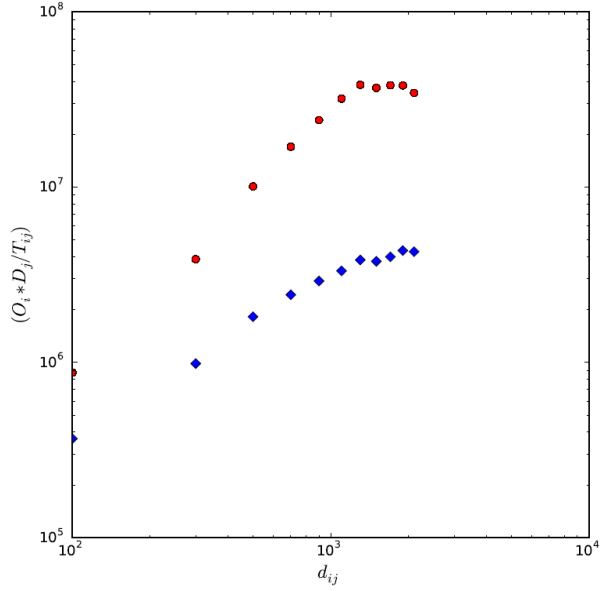

**Fig. 3.** The relationship between $(O_i \times D_j)/T_{ij}$ and $d_{ij}$. The blue points represent migration behaviors, and the red points represent travel behaviors. In fact, the distance function scaling exponent of migration behavior is 0.52, while the scaling exponent of travel behavior is 0.96.

In order to find the difference between the migration behavior and travel behavior, we study the resume data and the check-in data in accordance with another form of the gravity model formula. After taking the average from section, we find that the slope of these points equals the value of the parameter $\beta$ (Fig. 3). It can be seen that the absolute value of $\beta$ in migration behavior is lower than the absolute value of $\beta$ in travel behavior. In other words, long-term migration is less sensitive to distance than travel behavior. In our numerical calculation, we find that the distance function scaling exponent of migration behavior is 0.52, and the travel behavior's scaling exponent is 0.96.

### 3.4. Prediction model

The gravity model can be used to predict the moving flow, but the attenuation mechanism of the impedance function in the gravity model is more applicable to travel behavior in short time scales. For long-time scaled migration behavior, primordial gravity model does not work well. The reason is that the attenuation mechanism of classic gravity model depends mainly on distance. Nowadays, due to the higher accessibility among cities, the distance is not a most important factor in the people migration. Since migration is a long term behavior, the travelers are also very interested in the local economic level and personal development potential. The classical gravity model has been unable to apply to the present migration problem directly. Therefore, we have to build a new attenuation mechanism to describe and predict the migration behavior.

As we all know, prediction models usually take known and easily obtained data to predict unknown situations. We propose a new gravity model in order to predict trip flow of the migration behavior. The model directly uses both GDP and population data to predict the migration behavior. It is assumed that the amount of movement between the two places is proportional to the power function of the GDP and the population in the two places, and inversely proportional to the power function of the distance between the two places:

$$T_{ij} = \alpha \frac{P_i^{\theta_1} \cdot G_i^{\theta_2} \cdot P_j^{\theta_3} \cdot G_j^{\theta_4}}{d_{ij}^{\beta}} \quad (2)$$

where $T_{ij}$ is the amount of movement from the starting point $i$ to the destination $j$, $P_i$ and $P_j$ are the population of origin and destination respectively, $G_i$ and $G_j$ are the economic quantity (local GDP) of origin and destination, $d_{ij}$ is the distance between the two places, $\alpha$, $\beta$, $\theta_1$, $\theta_2$, $\theta_3$ and $\theta_4$ are adjustable parameters.

We apply the model to migration behavior and travel behavior, and compare computing results with actual resume data and Sina micro-blog check-in data (Fig 4). As you can see from the diagram, the gravity model's prediction in migration behavior is better than the prediction in travel behavior. It is worth mentioning that usually the statistics of population or GDP are analyzed separately for forecasting. When both of them are considered, the outcome could be much more accurate.

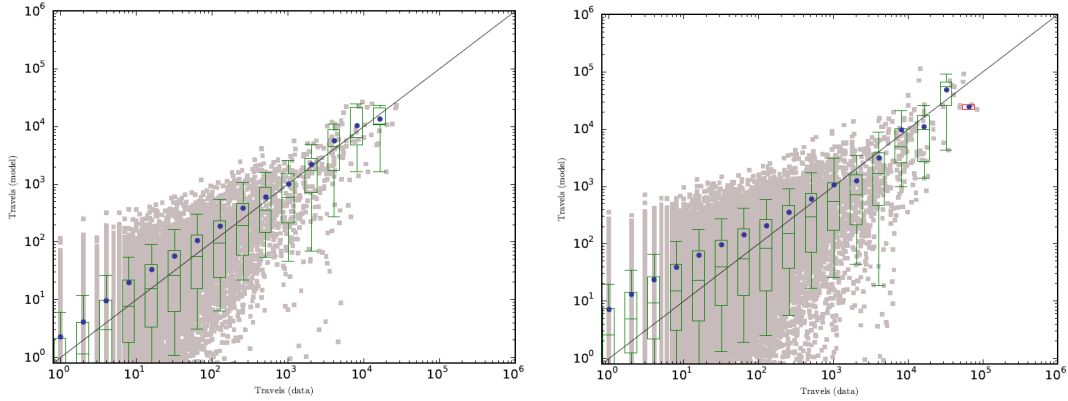

**Fig. 4.** Comparison between the observed fluxes and the predicted fluxes in two behaviors. (a) Migration fluxes predicted by the gravity model. (b) Travel fluxes predicted by the gravity model.

### *3.5. Comparison of model parameter*

In addition to comparing the predicted results of formula (2), we also compare the parameters of gravity model applied in different trip behaviors. Parameters of the gravity model used in migration behavior and travel behavior are shown in Figure 5.

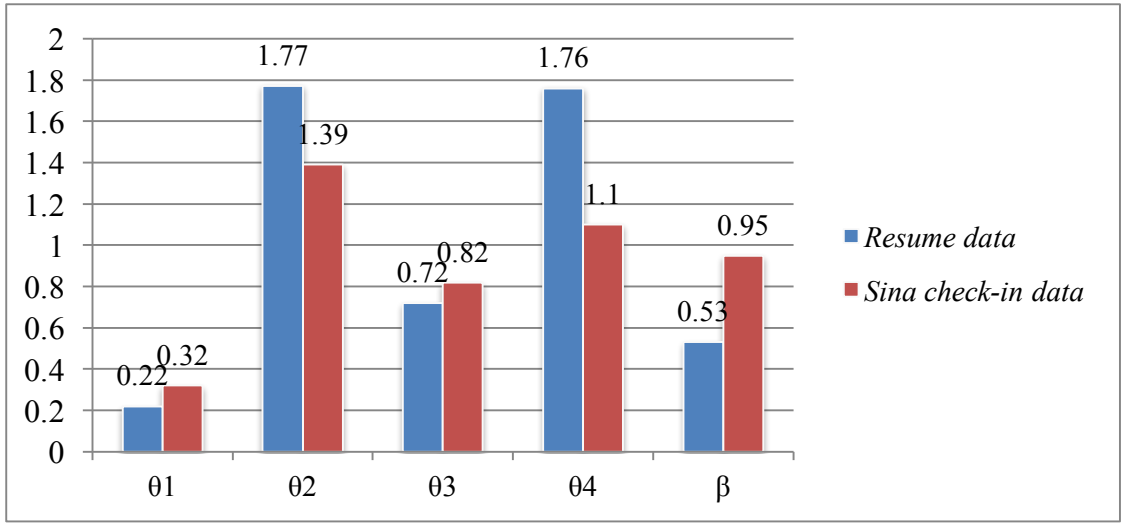

**Fig. 5.** Comparison between the migration behavior (Resume data) and the travel behavior (Sina check-in data) in terms of the parameters in the gravity model. The horizontal axis represents different parameters, and the vertical axis represents the parameter values. The blue column is the resume data, and the red column is the Sina micro-blog data.

According to formula (2) and figure 5, we can see that travelers (both migration behavior and travel behavior the two modes are available) are attracted more by the economic level than by the population size ($\theta_2$ or $\theta_4$ is greater than $\theta_1$ or $\theta_3$). Compared to the short-term travel behavior, long-term migration behavior is more affected by economic quantity (1.77>1.39, 1.76>1.1) and less influenced by population quantity (0.22<0.32, 0.72<0.82). By comparing the distance scale index $\beta$ of the two types, we can find that the long-term migration behavior is less affected by geographical distance than the short-term travel behavior. That is to say, migration behavior is less sensitive to distance.

*3.6. Comparison of Sorensen similarity index*

As shown in Table 1, we calculated the SSI (Sorensen similarity index) of the gravity model based on the migration behavior and the resume data which was 0.59. This shows that nearly 60% of migration behavior can be successfully predicted. It is worth mentioning that when we use gravity model based only on the population data to predict migration behavior, the SSI is only 0.31. This is very low. If we use gravity model based only on the GDP data to predict migration behavior, we can calculate that the SSI is about 0.44. The gravity model based only on the GDP is better than the gravity model based only on population data, but the SSI of the former is less than 0.5. In other words, the prediction results of the gravity model which is only based on single item need to be improved. That is why we built an improved gravity model for predicting migration behavior. The original definition of SSI is a very strict definition for comparing gene sequences, so the value of SSI is not very large for general travel behavior. Combining Formula (4) and Figure 4a, we can know that SSI tends to be biased towards larger amount of trip flow. Small amount of migrations has small contributions to the value of SSI. This is the reason why the value of SSI is relatively small.

**Table 1.** Prediction results of different gravity models

| Different gravity models | SSI |
| --- | --- |
| Use population only | 0.31 |
| Use GDP only | 0.44 |
| Both | 0.59 |

## 4. Discussion

Based on the resume data, this paper studies the transfer behavior among cities. The main contributions are: (1) the flow structure of long-term migration networks has highly asymmetric characteristics distinguished from short-term travel network which reflects the differences in attractiveness of cities and the competitive effects on the population; (2) the flow structure of the migration behavior can be approximately described with the gravity model in spatial economics. And the scaling exponent of distance function in the gravity model is less than the value of short-term travel network. This outcome shows that long-term migration behavior of people in the city is less sensitive to the space distance than the short-term travel behavior, and (3) according to the scaling coefficients of the parameters in the gravity model, the economic level has more influence on the migration population than the number of the local population.

In view of the contribution of this paper, we believe that the following work can be carried out in the future. In the gravity model, the distance scaling index of the migration behavior is approximately half of the travel behavior. Is it a coincidence? The mathematical equivalence relationship needs further calculation and demonstration. Moreover, cities with high levels of the economy seem to attract more talents. If we can find the relationship between flow of migration and economic quantity, we can predict the regional economic development trend through the structure of human mobility. This is of great significance in the field of economics.

**Acknowledgments**

This work was supported by the "China National Funds for Distinguished Young Scientists" (71525002), NSFC (71621001, 71671015).

**References**

[1] Barthélemy, M. (2011). Spatial networks. *Physics Reports, 499*(1–3), 1-101.
[2] Zipf, G. K. (1946). The PJVD Hypothesis on the Intercity Movement of Persons. *American Sociological Review* (Vol.11, pp.677-686).
[3] Jiang, B., Yin, J., & Zhao, S. (2009). Characterizing the human mobility pattern in a large street network. *Physical Review E Statistical Nonlinear & Soft Matter Physics, 80*(1), 021136.
[4] Roth, C., Kang, S. M., Batty, M., & Barthélemy, M. (2011). Structure of urban movements: polycentric activity and entangled hierarchical flows. *Plos One, 6*(1), e15923.
[5] Brockmann, D., & Helbing, D. (2013). The hidden geometry of complex, network-driven contagion phenomena. *Science, 342*(6164), 1337-42.
[6] Ratti, C., Sobolevsky, S., Calabrese, F., Andris, C., Reades, J., & Martino, M., et al. (2010). Redrawing the map of great britain from a network of human interactions. *Plos One, 5*(12), e14248.
[7] Murphy, E. (2012). Urban spatial location advantage: the dual of the transportation problem and its implications for land-use and transport planning. *Transportation Research Part A, 46*(1), 91-101.
[8] González, M. C., Hidalgo, C. A., & Barabási, A. L. (2008). Understanding individual human mobility patterns. *Nature, 453*(7196), 779.
[9] Candia, J., González, M. C., & Wang, P. (2008). Uncovering individual and collective human dynamics from mobile phone records. *Journal of Physics A,41*(22), 224015.
[10] Levandoski, J. J., Sarwat, M., Eldawy, A., & Mokbel, M. F. (2012). Lars: a location-aware recommender system. *, 41*(4), 450-461.
[11] Cho, E., Myers, S. A., & Leskovec, J. (2011). Friendship and mobility:user movement in location-based social networks. *ACM SIGKDD International Conference on Knowledge Discovery and Data Mining, San Diego, Ca, Usa, August* (pp.1082-1090). DBLP.
[12] Jiang, B., Yin, J., & Zhao, S. (2009). Characterizing the human mobility pattern in a large street network. *Physical Review E Statistical Nonlinear & Soft Matter Physics, 80*(1), 021136.
[13] Pappalardo, L., Simini, F., Rinzivillo, S., Pedreschi, D., & Giannotti, F. (2015). Returners and

explorers dichotomy in human mobility. *Nature Communications, 6*(8166).
[14] Riccardo, G., Armando, B., Sandro, R., & Marc, B. (2015). A stochastic model of randomly accelerated walkers for human mobility. *Nature Communications, 7*, 12600.
[15] Blumenstock, J., Cadamuro, G., & On, R. (2015). Predicting poverty and wealth from mobile phone metadata. *Science, 350*(6264), 1073-6.
[16] Toole JL, HerreraYaqüe C, Schneider CM, & González MC. (2015). Coupling human mobility and social ties. *Journal of the Royal Society, Interface / the Royal Society, 12*(105), 266-271.
[17] Marguta, R., & Parisi, A. (2015). Impact of human mobility on the periodicities and mechanisms underlying measles dynamics. *Journal of the Royal Society Interface, 12*(104), 20141317.
[18] Xu, Y., Shaw, S. L., Zhao, Z., Yin, L., Fang, Z., & Li, Q. (2015). Understanding aggregate human mobility patterns using passive mobile phone location data: a home-based approach. *Transportation, 42*(4), 625-646.
[19] Zhao, Z. D., Cai, S. M., & Lu, Y. (2015). Non-markovian character in human mobility: online and offline. *Chaos An Interdisciplinary Journal of Nonlinear Science, 25*(6), 15124-246.
[20] Yihong Yuan, & Martin Raubal. (2016). Analyzing the distribution of human activity space from mobile phone usage: an individual and urban-oriented study. *International Journal of Geographical Information Science, 30*(8).
[21] Chen, M., Yu, X., & Liu, Y. (2015). *Mining moving patterns for predicting next location*. Elsevier Science Ltd.
[22] Zhigang Li, Yan Shi, & Shanzhi Chen. (2016). Exploring the influence of human mobility on information spreading in mobile networks. *International Journal of Modern Physics C, 27*(06).
[23] Deville, P., Wang, D., Sinatra, R., Song, C., Blondel, V. D., & Barabási, A. L. (2014). Career on the move: geography, stratification, and scientific impact. *Scientific Reports, 4*(7497), 4770.
[24] Ortúzar, J. D. D., & Willumsen, L. G. (2011). *Modelling Transport, Fourth Edition*.
[25] Sørensen, T. (1957). A method of establishing groups of equal amplitude in plant sociology based on similarity of species and its application to analyses of the vegetation on danish commons. *Biol Skr, 5*, 1-34.
[26] Zipf, G. K. (1946). The p1p2/d hypothesis: on the intercity movement of persons. *American Sociological Review, 11*(6), 677-686.
[27] Hyman, G. M. (1969). The calibration of trip distribution models. *Environment & Planning A, 1*(1), 105-112.